# MAXIMUM PRODUCTION OF TRANSMISSION MESSAGES RATE FOR SERVICE DISCOVERY PROTOCOLS

Intisar Al-Mejibli[1] and Martin Colley[2]

[1,2]Department of Computer Science and Electronic Engineering, University of Essex, Essex, UK
ialmej@essex.ac.uk
martin@essex.ac.uk

***ABSTRACT***

Minimizing the number of dropped User Datagram Protocol (UDP) messages in a network is regarded as a challenge by researchers. This issue represents serious problems for many protocols particularly those that depend on sending messages as part of their strategy, such us service discovery protocols.

This paper proposes and evaluates an algorithm to predict the minimum period of time required between two or more consecutive messages and suggests the minimum queue sizes for the routers, to manage the traffic and minimise the number of dropped messages that has been caused by either congestion or queue overflow or both together. The algorithm has been applied to the Universal Plug and Play (UPnP) protocol using ns2 simulator. It was tested when the routers were connected in two configurations; as a centralized and de centralized. The message length and bandwidth of the links among the routers were taken in the consideration. The result shows Better improvement in number of dropped messages `among the routers.

## 1. INTRODUCTION

The Networked Home has become recognized as the forefront of the networking revolution, where consumer technology and Internet infrastructure intersect to change the way we lead our lives. Many researchers have noticed a fast growing increase in the use of home networks, for example recent research from Pike Research predicts a strong growth in the intelligent lighting control market. Global revenue is expected to increase from $1.3 billion to $2.6 billion by 2016 in intelligent lights [33].

In fact, a home network would consist of everything a homeowner could imagine, ranging from large domestic appliances such as the fridges, microwaves, audio-visual equipment to the lightweight temperature and smoke sensors. In addition to mobile devices, smart cards, bar codes in grocery packages and little chips in clothing and accessories. The main goal of interconnecting the home devices together is to share the network services and resources, and to invoke them remotely. Many protocols have been proposed to achieve this purpose which is locate and invoke the services and resources in network such as Universal Plug and Play (UPnP) these protocols are known as services discovery protocols [1]. Most of the service discovery protocols rely on the exchange of messages to locate remote services and to provide access to them. Sending too many messages onto the network from multiple nodes at the same time, could cause congestion which will lead to router queue overflow and the loss of messages. Accordingly, more messages must be sent to discover the services in the network and this causes more latency in discovery process and greedy consumption of the network resources. Hence, the router queue management algorithm should allow temporary bursty traffic, and penalize flows that persistently overuse bandwidth in order to avoid dropping packets. This paper argues that to avoid dropping messages during service discovery process in small networks which fall in (local Area Network) LAN category such as a networked office building, school, shop, or home. In addition, this paper proposes an algorithm to overcome this significant issue, in order to make the discovery process perform smoothly and seamlessly.

This paper is structured as follows. Section 2 introduces the related work which includes service discovery protocols, routing protocols, and available mechanisms and algorithms that have been proposed to avoid or minimize the number of dropped messages. Section 3 introduces the motivation of our proposal. Section 4 provides a discussion and analysis of the causes of dropped messages causes and the proposed algorithm. It also explains queue size algorithm, determining best interval algorithm and choosing candidate router rules in two network topology decentralized and centralized. The simulated model is detailed in section 5 and it contains two scenarios. In section 6, we present an evaluation of our algorithm. Finally, the conclusion is given in section 7.

## 2. RELATED WORK

We should introduce routing protocols, service discovery protocols, and the relevant work and algorithms in order to understand the subsequent sections.

## 2.1 ROUTING PROTOCOL

Many queue management algorithms have been proposed to manage queue strategy, such as Random Early Drop (RED) [2], Flow Random Early Drop (FRED) [3], BLUE [4], Stochastic Fair BLUE (SFB) [4], and CHOKe (a stateless active queue management scheme) [5]. In practice, most of the routers being deployed use the simplistic Drop Tail algorithm, which is simple to implement with minimal computation overhead, but provides unsatisfactory performance. Each queue management algorithm has its characteristics which differ from the other algorithms.

Random early detection (RED) is a queue management scheme that is intended to remedy the short comings of the drop tail algorithm. The gateway could notify connections of congestion by two methods: first: by dropping packets arriving at the gateway, second: by setting a bit in packet headers. The selected source is notified by the packet loss and its sending rate is reduced accordingly. Consequently, congestion is alleviated. RED gateways keep the average queue size low whereas allowing occasional bursts of packets in the queue.

RED is an early congestion notification because an arriving packet may be dropped before the queue is full. Depending on the queue length the dropping probability of RED is decided. The dropping probability is a function of average queue length. The RED dropping probability function is a piece wise linear function. It is defined by a triplet (*min*, *max*, *pa*). When the average queue size is greater than the *max*, every arriving packet is marked. If marked packets are in fact dropped, or if all source nodes are cooperative, this ensures that the average queue size does not significantly exceed the *max* threshold. Each arriving packet is marked with probability *pa* (where *pa* is a function of the average queue size), when the average queue size is between the *min* and the *max*, [2][27][35].

Flow Random Early Drop (FRED) [3][36] is a modified version of RED. FRED uses per-active-flow accounting to impose on each flow a loss rate that depends on the flow's buffer use. It does not make any assumptions about queuing architecture; it will work with a FIFO gateway.

Basically FRED is acts like RED but with the following modifications: FRED introduces the parameters *minq* and *maxq*, goals for the minimum and maximum number of packets each flow should be allowed to buffer. FRED introduces the global variable *avgcq*, an estimate of the average per-flow buffer count; flows with less than *avgcq* packets queued are favoured over flows with more. FRED maintains a count of buffered packets *qlen* for each flow that currently has any packets buffered. Finally, FRED maintains a variable strike for each flow, which counts the number of times the flow has failed to respond to congestion notification; FRED penalizes flows with high strike values [3].

BLUE [36] uses both of packet loss and link idle events to manage congestion. The key idea behind BLUE is to perform queue management based directly on packet loss and link utilization rather than on the instantaneous or average queue lengths. BLUE maintains a single probability, *pm*, which it uses to mark (or drop) packets when they are enqueued. If the queue is continually dropping packets due to buffer overflow, BLUE increments pm, thus increasing the rate at which it sends back congestion notification. Conversely, BLUE decreases its marking probability when the queue becomes empty or when the link is idle. This effectively allows BLUE to "learn" the correct rate it needs to send back congestion notification [4].

Based on BLUE, Stochastic Fair Blue (SFB) could be defined as technique for protecting TCP flows against non-responsive flows. SFB is a FIFO queuing algorithm depends on accounting mechanisms to identify and rate-limits non-responsive flows similar to those used with BLUE [4].

The CHOKe algorithm is built in the idea of maintaining a single FIFO buffer in the router side for queuing the packets of all the flows that share an outgoing link. Like the RED does the CHOKe calculates the average occupancy of the FIFO buffer using an exponential moving average. It also marks two thresholds on the buffer, a minimum threshold *minth* and a maximum threshold *maxth*. Every arriving packet is queued into the FIFO buffer when the average queue size is less than *minth*. The average queue size should not build up to *minth* very often and packets are not frequently dropped, when the aggregated arrival rate is smaller than the output link capacity. If the average queue size is greater than *maxth*, every arriving packet is dropped. This moves the queue occupancy back to below *maxth*. Each arriving packet is compared with a randomly selected packet, called drop candidate packet, from the FIFO buffer if the average queue size is bigger than *minth*. They are both dropped if they have the same flow ID. Otherwise, the randomly chosen packet is kept in the buffer and the arriving packet is dropped with a probability that bases on the average queue size. The drop probability is computed same way as in RED [5].

All the proposed solutions are applied to the router side, but our suggested solution is applied to the other nodes (any node at network that could be considered as a client or service nodes). The router has many functions to perform and by applying the protocol to the other nodes on the network the load in router side will be eased and this will have a positive impact on the whole network performance.

## 2.2 SERVICE DISCOVERY PROTOCOLS

Service discovery protocols enable devices to discover all services in a network and some of them allow devices that provide services to announce their services. Each service discovery protocol must have two components: a client which is the component that has a set of requirements that form the services it needs, and a device which is the component that offers its service(s) and is requested by client. Accordingly, any node in a network may be a client, a device, or a client and device at same time. Service discovery protocols can be classified into two types: Registry-based such as Jini [29][30] and Peer-to-Peer like UPnP. Registry-based can be further classified into centralized registry like Jini and distributed registry like Service Location Protocol (SLP) [29][31]. The Registry-based and Peer-to-Peer approaches both have advantages and drawbacks. For example: Registry-based is well organized and managed, but the registry node could cause a bottle neck problem for the entire network since if this node is damaged for any reason the clients are not able to access the required services. While in the Peer-to-Peer type all services send messages regularly even if there isn't a target client and this causes an unnecessary consumption for the networks' resources. Some protocols consider the announcement as an essential principle in service discovery issue such as UPnP whereas others protocols do not use the announcement approach such as Bluetooth [29] [32]. A selection technique should be used to select the most appropriate service when the discovery phase results in two or more identical services. There are two selection modes: manual and automatic modes. In manual mode, service selection is the responsibility of the user entirely. This mode has drawbacks: users may not know enough about the services to distinguish among them and too much user involvement causes inconvenience. This mode is applied in all the investigated service discovery protocols. In automatic mode, the service discovery protocol selects the service this simplifies client programs. On the other hand automatic selection may not be select the choice that user wants.

Each service discovery protocol has a specific features and philosophy which are different from other protocols. Here we will explain UPnP in more details as it is used in our simulated model.

### 2.2.1 UPnP

UPnP is proposed for use in home environments and small office and defines a method to discover device and service. It has the facility of automatically assigning IP addresses to networked devices. The components considered in UPnP are control points (clients) which are optional and devices (offers service(s)). Service discovery in UPnP is depended on the Simple Service Discovery Protocol (SSDP) [28]. SSDP was proposed to discover devices and services in a network easily, quickly, dynamically, and without any a priori knowledge [28]. It uses HTTP over unicast and multicast UDP packets to define two functions: search the services of a network and announce the availability of services in a network. UPnP cannot scale well since it uses multicasting extensively (multicasting is used both for service advertisements and service requests) [29]. When a control point is connected to network, it starts requesting the required service(s) by sending multicast message over UDP transport protocol. The service(s) that match the required criteria responds by sending unicast message to requested control point. Consequently, the control point gets information about the requested service. On the other hand, when the device is connected to network, it starts announcing its service(s) regularly by sending multicast message (over UDP).

## 2.3 TECHNIQUES AND MECHANISIMS

There are a number of techniques have been proposed to avoid dropping packets such as the approach which suggested by Parry and Gangatharan. The principle of their idea is the packet size of each source should be adjusted according to the network bandwidth to optimize the network utilization and also to avoid packet overflow at the client buffer. Their approach is based on a controller which is used to trace the data transmission rate at the router. When the total transmission rate is higher than the network bandwidth, the transmission controller adjusts the packet size of the source nodes so that the transmission rate is equal to the network bandwidth. [13].

Jacobson suggested an end-to-end congestion avoidance mechanism as used in Transmission Control Protocol (TCP). These mechanisms have worked well on low bandwidth delay product networks, while they have shown to be inefficient and prone to be unstable with newer high-bandwidth delay networks [14].

Jin proposed an alternative to the end-to-end congestion avoidance mechanism, named Network Lion and used in Transmission Control Protocol (TCP) too. Network Lion is developed as a part of a new network transmission protocol. His method employs a packet drop avoidance (PDA) mechanism based on of the maximum burst size (MBS) theory. In addition, he uses a real-time available bandwidth algorithm. Network Lion does redesign the transmission control, as well as separating the pacing control in layer 3 and retransmission control in layer 4. [15].
Kevin Mills and Christopher Dabrowski [16] proposed four Algorithms for adaptive-jitter control depending on network size, in order to minimize the dropping of messages from the message queues in the UPnP protocol. In fact, UPnP permits clients to include a jitter bound in multicast (M-Search) queries in order to limit implosion. Qualifying devices use the jitter bound to randomize timing of their responses. Kevin Mills and Christopher Dabrowski's algorithms depend on the principle of this bound. All four of these algorithms are based on making each root device independently estimate the time it will take for all root devices to respond to each M-Search query. Each root device then uses its estimate to determine a time to send its own responses (if any). Each response message includes a value recommending how long the control-point M-Search task should listen for responses, so M-Search task does not need to guess an appropriate required maximum time for listening.
All root devices must send and listen to Notify messages (which include a caching time or *max-age*). When all root devices receives these messages, they should build a map (*NM*) of devices and services in the network. Consequently, a root device could use its *NM* to estimate how many response messages will be sent by all root devices. They assume that the messages will be sent consecutively at rate *R and* root devices will send messages sequentially in the ascending order of their unique identities.

# 3. MOTIVATION

Many of the protocols that address the issue of service discovery in network are dependent on sending messages, such as UPnP, SLP, and Jini. Depending on this mechanism to discover the services in the network could result in the dropping of messages. Accordingly, dropped messages is a result of sending too many messages to the network without considering the bounds of the network resources such as the bandwidth, existing queue sizes, network size, etc. In fact, this result (dropped messages) caused another issues such as latency in discovering process and resending the dropped messages.

Actually, each network as physical components has limited capabilities that must be considered in applying any protocol, to avoid any fault and get optimal results. So in applying a service discovery protocol the physical network resources must be taken into consideration to avoid dropping messages.

Although, in small network, the number of connected nodes and there location are dynamically changed, most of the physical components of the network resources such number of routers and bandwidth are fixed. Knowing the network abilities in advance is significant too to avoid dropping messages at the router side.

This paper explains the relation between the size of the burst of messages, the size of the queue in router side, bandwidth and number of dropped messages. It proposes a solution to avoid dropping messages in router side and this solution is presented as an algorithm that takes into consideration the network size, diagram, the used link bandwidth and messages sizes.

Some of the suggested solutions for this issue which mentioned in the related work were discussed this issue and solution for TCP packets only and some of them suggested a controller in the router side. No one of the previous solutions discuss the ability of solve this issue in the sending side in other words in the nodes that applied the specific protocol not in the router side.

# 4. DISCUSSION AND ANALYZING

It is essential to analyze the causes of dropped messages to facilitate discussion in the proposed algorithm.

## 4.1 DROPPING MESSAGES CAUSES

A queue is used when there are packets between two applications or networks. One of them inputs packets onto the queue and the other outputs these packets, so queues are a solution for the asynchronous flow of packets or operations, see F 1. If the incoming rate is lower than the outgoing rate, the outgoing queue (Queue1) will be almost empty and the forwarding process not blocked and there are no dropped packets. On the contrary, the outgoing queue (Queue1) will be almost full when the incoming rate higher than the outgoing rate. In overflows, packets have to be dropped because there is not enough space in the outgoing queue to store them for preserving its survival [23].

The chance of queue overflow can be reduced by allocating more memory for the queue, but this is difficult to predict because of the variation in incoming rate [28]. The best or ideal queue size should be enough to cope with bursts of messages which is one of the common causes of dropped messages.

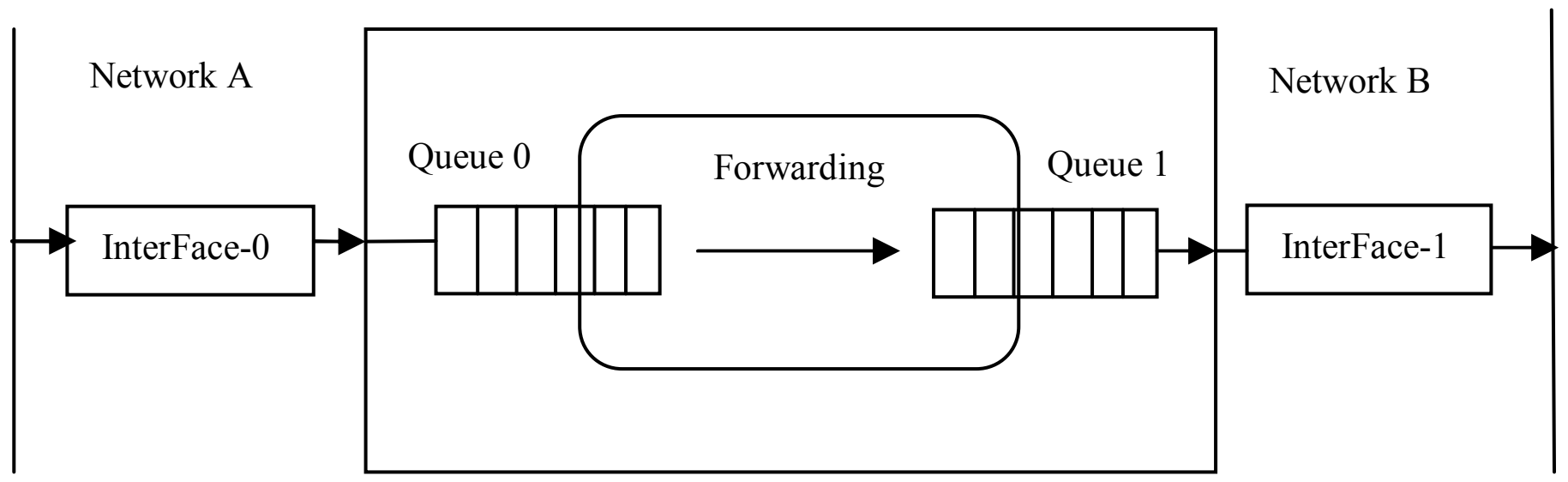

Figure 1. Shows the strategy of queue in router

Suppose that $f(SMsg_{Ci}) = \sum_{k=1}^{N}(SMsg_k)$ is number of all the sent messages by the services to client *i, i=1,2, ...,m.*

Where, N is the number of all services in the network, M is the number of all the clients in the network. Further suppose that client $i$ needs $RT_{Ci}$ time to receive all messages. The incoming rate ($IR_{Ci}$) and average processing rate ($PR_{Ci}$) are measured by messages / second so:

$IR_{Ci}$ / 1 = $f(SMsg_{ci})$ / $RT_{Ci}$

$RT_{Ci}$ = $f(SMsg_{ci})/IR_{Ci}$. . . . . . (1)

Equation (2) shows the number of processed messages $f(PMsg_{Ci}) = \sum_{k=1}^{P}(PMsg_k)$ during $RT_{Ci}$, p is the number of processed messages.

$PR_{Ci}$ / 1 = $f(PMsg_{Ci})$/ $RT_{Ci}$

$f(PMsg_{Ci})$= $RT_{Ci}$ * $PR_{Ci}$...... (2)

Now suppose $f(RMsg_{Ci}) = \sum_{k=1}^{R}(RMsg_k)$ is the number of received messages by *Ci, and* $Qsize_{Ci}$ is the queue size of *Ci* then*:*

$$f(RMsg_{Ci}) \begin{cases} \leq f(PMsg_{Ci}) + Qsize_{Ci} \ \ There\ won't\ be\ dropped\ messages \\ > f(PMsg_{Ci}) + Qsize_{Ci} \ \ There\ will\ be\ droppeed\ messages \end{cases}$$

Accordingly, there would not be any dropped messages if the condition in equation 3 is satisfied. Otherwise there will be dropped messages.

$f(RMsg_{ci})$ <= ($RT_{Ci}$ * $PR_{Ci}$) + $Qsize_{Ci}$.................................................... (3)

Note: The result of equation 3 will always rounded up.

From Equation (3) the queue size is:

$Qsize_{Ci}$= $f(RMsg_{ci})$ – [$RT_{Ci}$ * $PR_{Ci}$) ]..... (4)

The best queue size ($BQsize_{Ci}$) which is guaranteed not to drop messages, must be equal or greater than the $Qsize_{Ci}$ in equation 4. Otherwise there will be dropped messages.

From equation (4) the $RT_{Ci}$equal to:

$RT_{Ci}$ = ( $f(RMsg_{ci})$– $Qsize_{Ci}$) / $PR_{Ci}$ …. (5)

$$\text{If } RT_{Ci} \text{ is} \begin{cases} \leq \frac{(f(RMsg_{ci}) - Qsize_{Ci})}{PR_{Ci}} \ \ There\ won't\ be\ dropped\ messages \\ > \frac{(f(RMsg_{ci}) - Qsize_{Ci})}{PR_{Ci}} \ \ There\ will\ be\ droppeed\ messages \end{cases}$$

## 4.2. The PROPOSED ALGORITHM

The proposed algorithm is built on the idea of determining the required receiving queue space to avoid losing messages. In addition to determining the required time for the router to be ready to receive the next burst of messages without dropping messages. Accordingly, the proposed algorithm explains the relation between the required queue sizes and the interval separating two consecutive bursts of messages, to avoid dropping messages. The following rules must be applied to compute the receiving queue size in each router or the space which required being available in the receiving queue of each router at the sending time and calculate the best interval for each router. The algorithm was tested when the routers were connected in two configurations centralized and decentralized.

### 4.2.1. Decentralized Network

The routers in this configuration are connected decentralized methode see F 7. Suppose $SQsize_{Ri}$ is the sending queue size of router *Ri* and $f(C_{Ri}) = \sum_{k=1}^{Rim}(C_k)$ , $f(S_{Ri}) = \sum_{k=1}^{Rin}(S_k)$ are the total number of clients and services that connected to *Ri* consecutively.

#### 4.2.1.1. Queue size Algorithm:

1. If a router is connected to services only then: $SQsize_{Ri} = Rin + 2$ .
2. If a router not connected to any node then: $SQsize_{Ri} = 2$.
3. Otherwise: $SQsize_{Ri} = Rim + Rin + 1$.

**4.2.1.2. Determining best interval Algorithm:**

In a de centralized network any chosen router will divide the network into two parts left and right. Thus it is required to define a time function; $(T(x_{\mathrm{k}})) = \frac{Message\ Size\ of\ service_{\mathrm{k}}}{Bandwidth\ that\ message\ would\ use}$

To calculate the best interval for a router connected to one client (client acts as receiver) or more:

1. Identify the candidate routers $\mathrm{CR_h}$ where h=1, 2, ... z, z is the number of candidate routers.
2. For each router $\mathrm{CR_j}$ j=1,2,... z do the following start:
3. Identify the two parts left and right of $\mathrm{CR_h}$.
4. Calculate the total number of services that comes from left $(\sum_{k=1}^{Lri}(Sl_{\mathrm{k}}))$ and right $(\sum_{k=1}^{Rri}(SR_{\mathrm{k}}))$.
5. Chose the larger part: Large $\begin{cases} = Lri\ if\ Lri > Rri \\ = Rri\ if\ Rri > Lri \end{cases}$
6. Calculate the best interval of $\mathrm{CR_h}$ $(\mathrm{BICR_h}) = \sum_{k=1}^{\mathrm{Large}}(T(x_{\mathrm{k}})) + \sum_{k=1}^{\mathrm{gaps}}(T(x_{\mathrm{k}})) - (T(x_{\mathrm{j}}))$. . (6)
7. End for
8. Best interval (BI) will be equal to the largest value among $(\mathrm{BICR_h})$ values h=1, 2, ..., z.

$(T(x_{\mathrm{j}}))$ value is the biggest among $(Time(x_{\mathrm{k}}))$, $k = 1,2, \ldots Large$. $(T(x_{\mathrm{j}}))$ represents the time the message utilizes the link.

$\sum_{k=1}^{\mathrm{gaps}}(T(x_{\mathrm{k}}))$: represents the number of messages times during which a specific router doesn't receive any service messages from nearest router(s). Here the average message size and average bandwidth is used. When there is a service connected directly to the nearest router, it would need at least two message times to reach the evaluated router.

Equation (6) guarantees that a specific router would forward all the receiving messages to their destination (client) before receiving the next burst of messages. It can be developed and take into consideration the available receiving queue size for the specified router, as it represents the sharing space between all the clients (receivers) connected to that router so an overlap between two or more consecutive burst of messages can be achieved in order to minimize the required interval.

The following steps show how to calculate the Overlapped space (OS):

1. Identify the router neighbor to chosen router and this will be *Rneighbour* = *RLarge.*
2. Identify the sending queue size of RLarge which is: $SQsize_{RLarge}$
3. Overlapped space (OS): OS = $(SQsize_{RLarge} - f(S_{RLarge}))$ / $f(C_{Ri})$. . . . (7)

Equation (1) could be written as:

The best interval (BI) = $\sum_{k=1}^{\mathrm{Big}} T(x_k) + \sum_{k=1}^{\mathrm{gaps}} T(x_k) - T(x_j) - \sum_{k=1}^{\mathrm{OS}} T(x_k)$. . . (8)

The question now, must each router in a network be evaluated in order to identify the best interval for entire network? And which interval would be used for the network? The answer is: Not all routers in a network must be evaluated instead some of them would be candidate to be evaluated and the longest interval will be used at the end, because, logically using the longest interval will avoid dropping messages at all other routers.

There are some conditions that help to identify which router will have the most impact in determining the best interval.

#### 4.2.1.3. Choosing candidate router rules:

**1.** Identify the longest path between a service and a client. Candidate the router that is connected to this client.

**2.** Identifying the router that connected to the largest number of clients and receives the largest number of services from one side of the network.

**3.** Identifying the router that connected to one or more clients and located nearest the end of the network.

**4.** If the chosen router is connected to one client then the nearest router connected to client too must be chosen, in order to compare between two consecutive burst of messages reach these routers consecutively. In case that the two (or more) consecutive burst of messages were sent to the same client and this client is the lonely client connected to router, this means logically there are two (or more) receivers connected to that router and this should be taken into consideration in calculating the (OS) value.

One client may satisfy more than one of the previous conditions, in other words the client that has longest path with a service could be the same client that connected to a router which receives largest number of services and this wouldn't cause any problem.

All the candidate routers must be evaluated and the longest interval is the best interval for the network which would guarantee no losing messages.

### 4.2.2 Centralized Network

When the configuration of the network is centralized, see F 8, there will be a difference in the rules used for calculating the queue size and best interval in addition to the conditions which should be satisfied in choosing the candidate routers.

#### 4.2.2.1 Queue size algorithm

Identifying the router queue size is dependent on its location in the network

**1-** Root router: identifying its sending queue size required the following process:

**2-** Identify the router that directly connected to the largest number of services (*RLargeS*). Note: when there are two or more routers connected to the same number of services, any of them may be chosen.

**3-** Determine the required time that all services which connected to *RLargeS* need to reach root router ( $RT_{RLargeS}$).

$RT_{RLargeS}=T(x_j)+\sum_{k=1}^{\text{RLargeSn}}T(x_k)$. . . *(9).*

Where $T(x_j)$ value is the largest value among $T(x_k),\ k=1,2,\ldots RLargeSn$ values. $T(x_k)$ represents the first interval which there isn't any service reach root router and to get a best results the algorithm choose the longest interval. Note *RLargeSn* is the number of services connected to *RLargeS*.

**4-** During $RT_{RLargeS}$ the root router will be already forwarded *RLargeSn*-1 messages to the next router.

**5-** The minimum sending queue size for root router should be:

$SQsize_{Rroot}$= *n* - *RLargeSn* - (*RLargeSn*-1 ). . . *(10).* Where *n* is the number of services in the entire network.

**6-** If a router is not a root router, then minimum sending queue size is calculated as following:

$SQsize_{Ri}= f(C_{Ri})+ f(S_{Ri})$. Where i=1,2,…,*RSum*-1 and *RSum* is the total routers in the network.

#### 4.2.2.2 Determining best interval algorithm:

It is clear that any sending message between any two router should pass the root router, in order to reach its destination (root router is represented in R5 as shown F8).
Determining the ideal interval that can cope with the suggested queue sizes required:
1. Identify the router that is connected to the largest number of clients *RLargeC.*
2. If there is more than one router satisfying the previous condition, the chosen router would be the router that receives the largest number of services *RLargeC_S.*
3. The best interval is = $\sum_{k=1}^{n} T(x_k) + \sum_{k=1}^{\text{gaps}} T(x_k)$ - $T(x_j)$ - $\sum_{k=1}^{RLargeC_S} T(x_k)$ . . . (11)
$\sum_{k=1}^{\text{gaps}} T(x_k)$ represents the number of messages times during which a specific router doesn't receive any service messages from nearest router(s), and $T(x_j)$ represents the time the message utilizes the link.

#### 4.2.2.3 Choosing candidate router rules:

To choose the proper router that influences the length of the interval.
**1.** Identify the router that receives the largest number of services.
**2.** If the chosen router is connected to one client only then there are two options depending on the applied protocol. The first option is when the applied protocol would send two or more consecutive burst of messages to this client, logically there are two or more receivers connected to that router and this should be taken into consideration in calculating the (OS) value. The second option is identify the nearest router that connected to two (or more) clients.

# 5. SIMULATION RESULTS

The following two scenarios explain the previous rules in both network configurations de centralized and centralized. The used bandwidth is 512Kb with delay (0ms), the message length is 128 bytes and type of used queue is DropTail. The NS2 simulator is used to perform these experiments.

## 5.1 Scenario1

The network design includes 4 routers, each router is connected to 4 clients (in dark colour) and 20 (in white colour) services as shown in F 2. The applied scenario is:
1- All clients send multicast message to discover all the services in the network, then,
2- All services send reply messages to first client at the same time then wait for an interval time then send reply messages to the second client and so on.

All the routers are symmetric in number of connected client and services. So we would depend on the longest path between client and service in choosing the candidate router. It is clear that the services which are connected to R0 have longest path to reach clients in R4. On the other hand, services connected to R4 have longest path to reach clients in R0. R0 and R4 are the candidate routers and because they are symmetric one of them will be chosen; let it be R0.

Depending on the queue size algorithm and as all the routers are connected to clients and services, the applied formula is $SQsize_{Ri} = Rim + Rin + 1$, $Ri$= 0,1,2,3.

$SQsize_{R0}$ = 20+4+1= 25 packets. So the sending queue size for R0, R1 ,R2 , and R3 = 25 packets.

To calculate the best interval the Overlapped space (OS) must be calculated

Overlapped space (OS): OS = (25 –20) / 4.

OS = 5 / 4, OS = 1 message space.

All sent messages are equal in size and all links are equal in bandwidth so

$T(x_k)$ for all $k$ values are equal. $(T(x_k)) = \frac{128*8}{512*1024}$ , $(T(x_k)) = 0.001953125 \approx 0.002$ second.

The best interval (BI) = $\sum_{k=1}^{80} 0.002 + \sum_{k=1}^{1} 0.002 - 0.002 - \sum_{k=1}^{1} 0.002$

The best interval (BI) = $0.16 + 0.002 - 0.002 - 0.002$

The best interval (BI) = 0.158 second

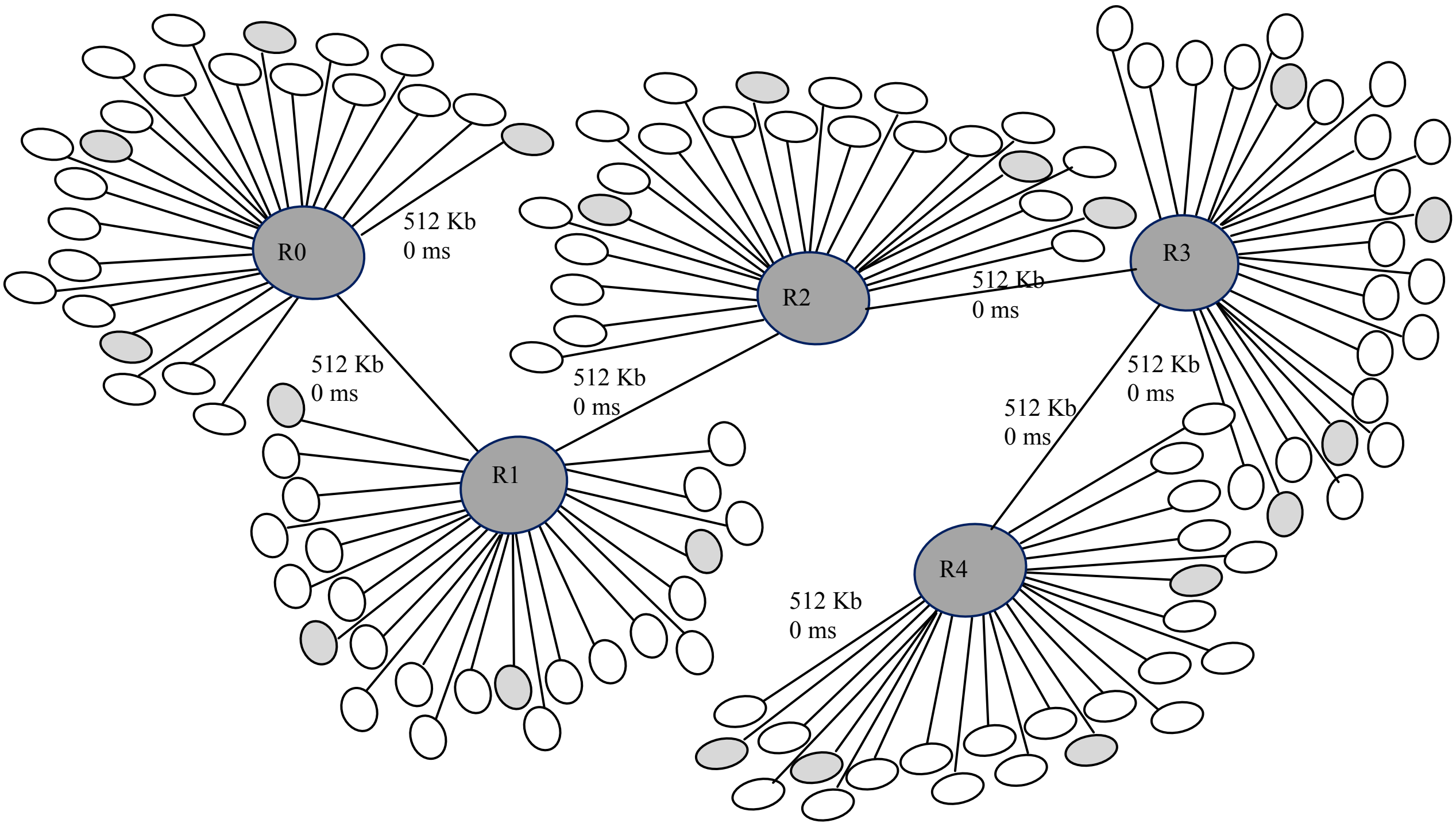

Figure 2. Network with 4 routers, 100 services and 20 clients

When this interval is applied in NS2 simulator, the result shows that there are no lost messages. The required time to discover the entire network was 3.246 seconds; this time includes the multicast time (0.042 second) and unicast time (3.204 second).

The previous scenario supposes there is not any back traffic in the network. If the scenario is changed as the follows:

1- All clients send multicast message to discover all the services in the network, then,
2- All services send reply messages to first client at the same time then wait for 0.158 seconds then send reply messages to the second client and so on.
3- One service on the R0 sends 20 messages to a client connected R2 one every 0.158 second. Started nearly at the same time of point 2.
4- One service on the R3 sends 20 messages to a client connected R4 one every 0.158 second. Started nearly at the same time of point 2.

The results show dropping one message between R3 and R4.

## 5.2 Scenario 2

This scenario is the same previous one but it is applied in the centralized network. All routers are not connected to each other directly rather than, they connected to one router which root router, see F 3.

In this scenario each router of (R0, R1, R2, R3, and R4) are connected to 20 services (in white colour) and 4 clients (in dark colour). The applied scenario is:

1- All clients send multicast message to discover all the services in the network, then,

2- All services send reply messages to first client at the same time then wait for an interval time then send reply messages to the second client and so on.

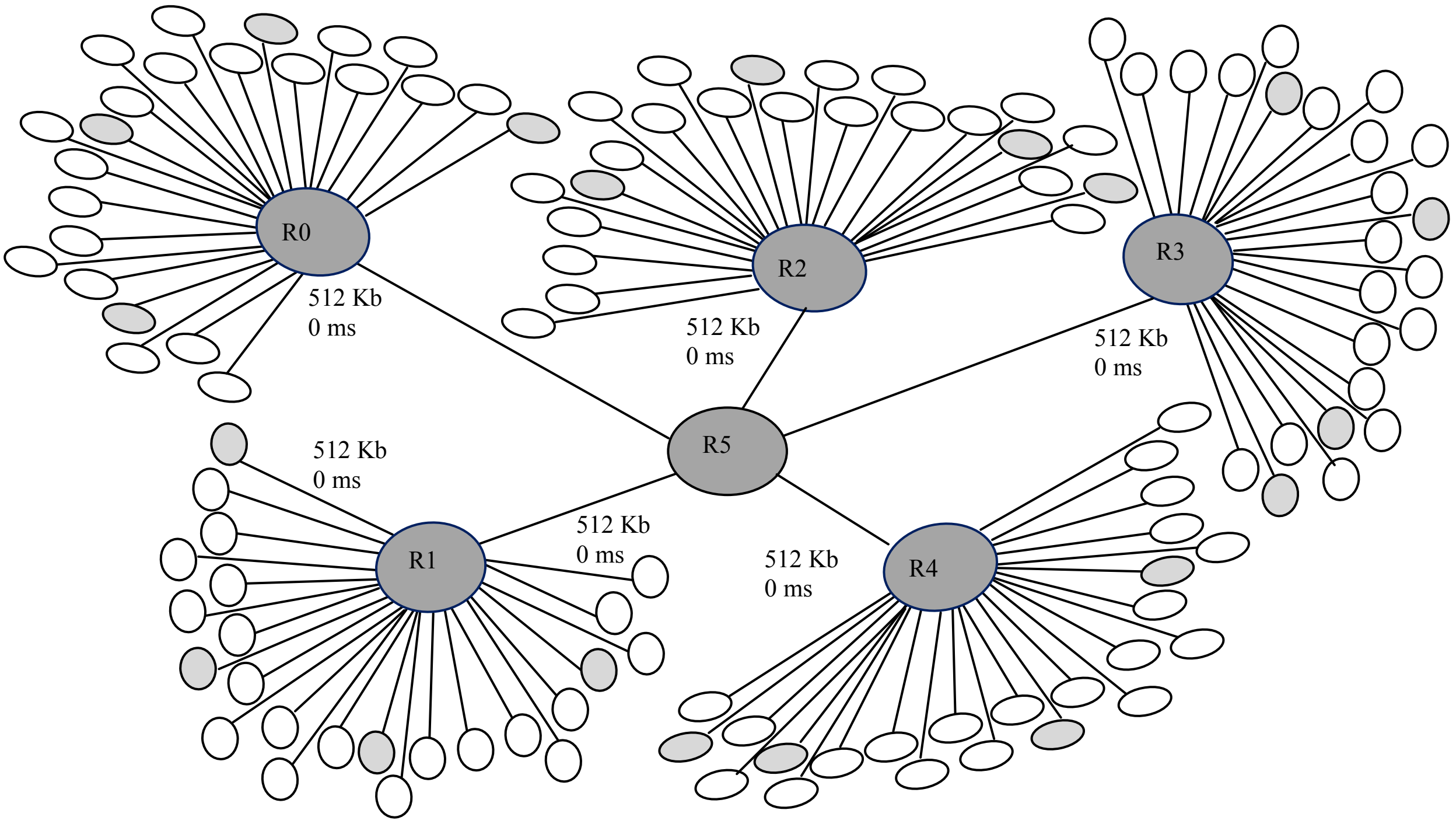

Figure 3. Network with 4 routers, 100 services and 20 clients

To calculate the sending queue size for the root router, *RLargeS* router must be identified. All the routers are symmetric, so R0 will be regarded as *RLargeS* router.

Queue size algorithm

a- R0 is the *RLargeS.*

b- $RT_{RLargeS}$=0.002 + $\sum_{k=1}^{20} 0.002$, $RT_{RLargeS}$=0.042.

c- Forwarded messages = *RLargeS - 1,* Forwarded messages = 20 -1 =19 messages.

d- $SQsize_{Rroot}$= *100 – 20 - (20*-1) , $SQsize_{Rroot}$= *61 messages.*

Calculate the sending queue size for the other routers using this formula

$SQsize_{Ri}= f(C_{Ri}) + f(S_{Ri})$

As all the routers are symmetric then $SQsize_{R0} = SQsize_{R1} = SQsize_{R2} = SQsize_{R4}$

$SQsize_{R0}$= 4+20 = 24 messages.

Now apply the determining best interval algorithm

1. Let R0 is the *RLargeC*.
2. *RLargeC_S*= $\sum_{k=1}^{R0n} f(x_k)$, *RLargeC_S* = *20*.
3. The best interval (BI) is = $\sum_{k=1}^{100}(0.002)$ + $\sum_{k=1}^{2}(0.002) - 0.002$ - $\sum_{k=1}^{20}(0.002)$.

BI = 0.2+ 0.004 - 0.002 -0.04.
BI = 0.162 second.

When this interval is applied in NS2 simulator, the result shows that there are no lost messages. The required time to discover the entire network was 3.322 seconds; this time includes the multicast time (0.042 seconds) and unicast time (3.28 seconds).

The previous scenario supposes there is not any back traffic in the network. If the scenario of the previous scenarios is changed as the following:
1- All clients send multicast message to discover all the services in the network, then,
2- All services send reply messages to first client at the same time then wait for 0.162 seconds then send reply messages to the second client and so on.
3- One service on the R0 sends 20 messages to a client connected R2 one every 0.162 second. Started nearly at the same time of point 2.
4- One service on the R3 sends 20 messages to a client connected R4 one every 0.162 second. Started nearly at the same time of point 2.
The results show that there are no lost messages. So if there is any back traffic or not, the results indicate there will not be any dropped messages.

Table 1 shows the best interval time for a variety network sizes start from 8 routers 8 clients and 8 services to 16 routers, 160 clients and 160 services.

Table 1. Shows the best interval time for a variety network sizes

| **Each router is connected to** | **8 routers** | **12 routers** | **16 routers** | **Network size** |
|---|---|---|---|---|
| 1 client & 1 services | *TSoM * 4* | *TSoM * 8* | *TSoM * 12* | *2 * router No.* |
| 2 client & 2 services | *TSoM * 12* | *TSoM * 20* | *TSoM * 28* | *4 * router No.* |
| 3 client & 3 services | *TSoM * 20* | *TSoM * 32* | *TSoM * 44* | *6 * router No.* |
| 4 client & 4 services | *TSoM * 27* | *TSoM * 43* | *TSoM * 59* | *8 * router No.* |
| 5 client & 5 services | *TSoM * 34* | *TSoM * 54* | *TSoM * 74* | *10 * router No.* |
| 6 client & 6 services | *TSoM * 41* | *TSoM * 65* | *TSoM * 89* | *12 * router No.* |
| 7 client & 7 services | *TSoM * 48* | *TSoM * 76* | *TSoM * 104* | *14 * router No.* |
| 8 client & 8 services | *TSoM * 55* | *TSoM * 87* | *TSoM * 119* | *16 * router No.* |
| 9 client & 9 services | *TSoM * 62* | *TSoM * 98* | *TSoM * 134* | *18 * router No.* |
| 10 client & 10 services | *TSoM * 69* | *TSoM * 109* | *TSoM * 149* | *20 * router No.* |

The routers are connected to each other in decentralized method like in F 2. Where TSoM represents Time to Send one Message and the first columns shows the number of connected clients and services to each router.

# 6. EVALUATING THE SUGGESTED ALGORITHM

There are two considerations or bases in which the proposed algorithm could be evaluated.

1- First base is the changing in the number of nodes in the network.
2- The proposed algorithm used the number of the nodes in the entire network to determine the required interval. The other way supposes that the clients continuously send discovery messages separated by a suggested interval. This suggested interval usually has a maximum limit.

## 6.1 First consideration

The nodes in small network (home network or office network) are not fixed, on the contrary; it is dynamically changing because nodes frequently join or leave. This is the first challenge that the proposed algorithm must face it.

In the case of the de centralized network the formula (8) shows how to calculate the best and formula (7) shows how to calculate OS value. So the best interval value depends on the number of the services in the network and on the OS values when these two values changed the interval may be need to recalculate. If number of the services is increased or OS value decreased the BI must be recalculated. Otherwise messages would be dropped.

In the case where the number of the services is decreased or OS value increased, the BI value could still be used and there would not be any dropped messages. In this case the used interval will be more than required. OS value depends on the chosen router if the chosen router lost its features that made it as the best candidate router and another router gain this feature then the OS value should be recalculated to prevent messages from being dropped.

Regarding to the centralized network formula (11) shows how to calculate the best interval. The best interval depends on the number of services in the entire network and on the router that connected to the largest number of clients and receives the largest number of services. If any parameter is changed, this will impact on the value of the best interval. When new service(s) are added to the network the BI value must be recalculated, otherwise messages will be dropped. While when a service(s) leaves from the network, there will not be any dropped messages even the old value of BI is used.

The chosen router which is connected to the largest number of clients and receives the largest number of services is the second critical value in calculating the best interval. If this router is changed, this happened when the connected clients to this router are disconnected and connected to another router or when new client(s) are connected to another router then the BI should be recalculated to avoid dropping messages. If the value $RLargeC_S$ decreased then the BI must be recalculated to avoid dropping messages. Otherwise there would not be dropped messages.

## 6.2 Second consideration

The clients stop sending discovery messages when there are no more reply messages from the services because this indicates that all of the network has been discovered. The principle of this way is that: the client send the discovery message and wait for services responses. Then send another discovery message with the names of discovered services to avoid their double response (as they discovered before), and so on until no response is received. To evaluate the proposed algorithm with this method that depends on maximum limit, two models are designed and applied using NS2 simulator.

### 6.2.1 Decentralized model

The network design includes 6 routers each router is connected to 4 services (in white colour) except router R2 which connected to two clients (in dark colour) as shown in F 4. The applied scenario is:

1- All clients send multicast message to discover all the services in the network, then,
2- All services send reply messages to first client at the same time then wait for an interval time then send reply messages to the second client and so on.
3- At first the proposed algorithm is used to calculate the required interval and it was 0.048 seconds. Then the maximum limit method is used with different values (0.15, 0.1, and 0.05 seconds). Table 2, F 5 and F 6 shows the results:

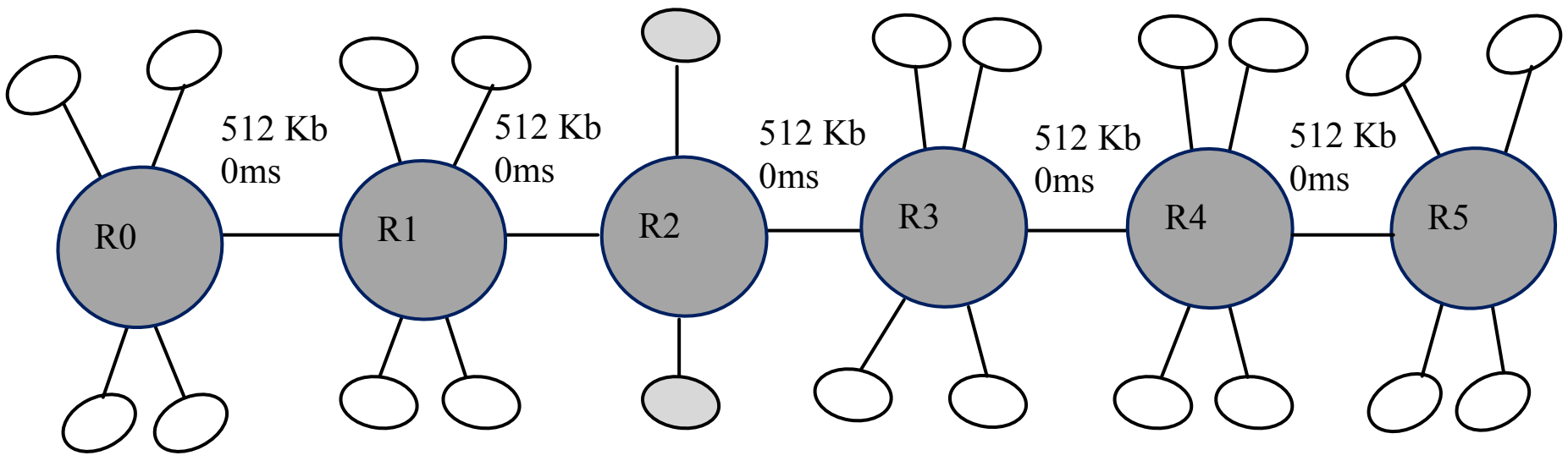

Figure 4. Network with 6 routers, 20 services and 2 clients

Table 2 shows a comparison between the proposed algorithm and the Maximum limit discovery method

| **Algorithm** | **Maximum limit discovery method** | | |
|---|---|---|---|
| | **0.15 second** | **0.1 second** | **0.05 second** |
| 1-Clients need to send one discovery message only (multicast messages) | 1- Clients need to send Three discovery messages (multicast messages) | 1- Clients need to send Three discovery messages (multicast messages) | 1- Clients need to send Six discovery messages (multicast messages) |
| 2- No dropped messages | 2- There are dropped messages round 9 messages | 2- There are dropped messages round 9 messages | 2- There are dropped messages round 24 messages |
| 3- Number of the sent replying messages is 40 messages | 3- Number of the sent replying messages is 49 messages. | 3- Number of the sent replying messages is 54 messages. | 3- Number of the sent replying messages is 102 messages. |
| 3- Receiving Duplicate 0 | 3- Receiving Duplicate 0 | 3- Receiving Duplicate 5 | 3- Receiving Duplicate 38 |
| 4- Discovery process needs 0. 162 seconds to be achieved | 4- Discovery process needs 0.45 seconds to be achieved | 4- Discovery process needs 0.3 seconds to be achieved | 4- Discovery process needs 0.3 seconds to be achieved |
| 5- All services send one reply messages to each client at one time. No repeating in sending the replying messages | 5- First phase clients discovered 77.5 % of the services, at second phase 100% of the services discovered, third phase to make sure. | 5- First phase clients discovered 65 % of the services, at second phase 100%, Third phase to make sure. | 5- First 7.5 % , second 57.5%, third 85%, fourth 95%, fifth 100%, Sixth phase to make sure. |

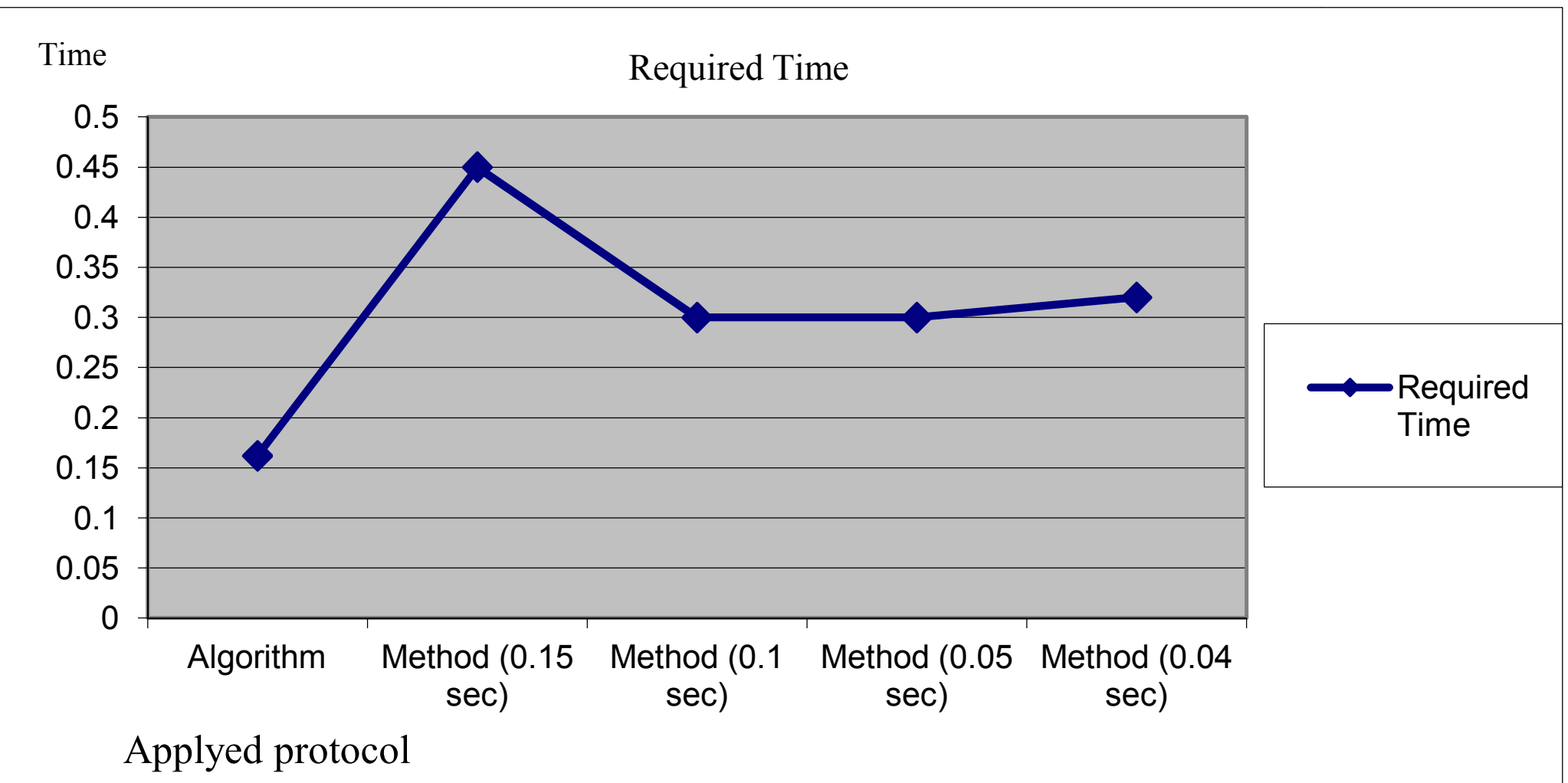

Figure 5 shows the difference in the required time in the two ways

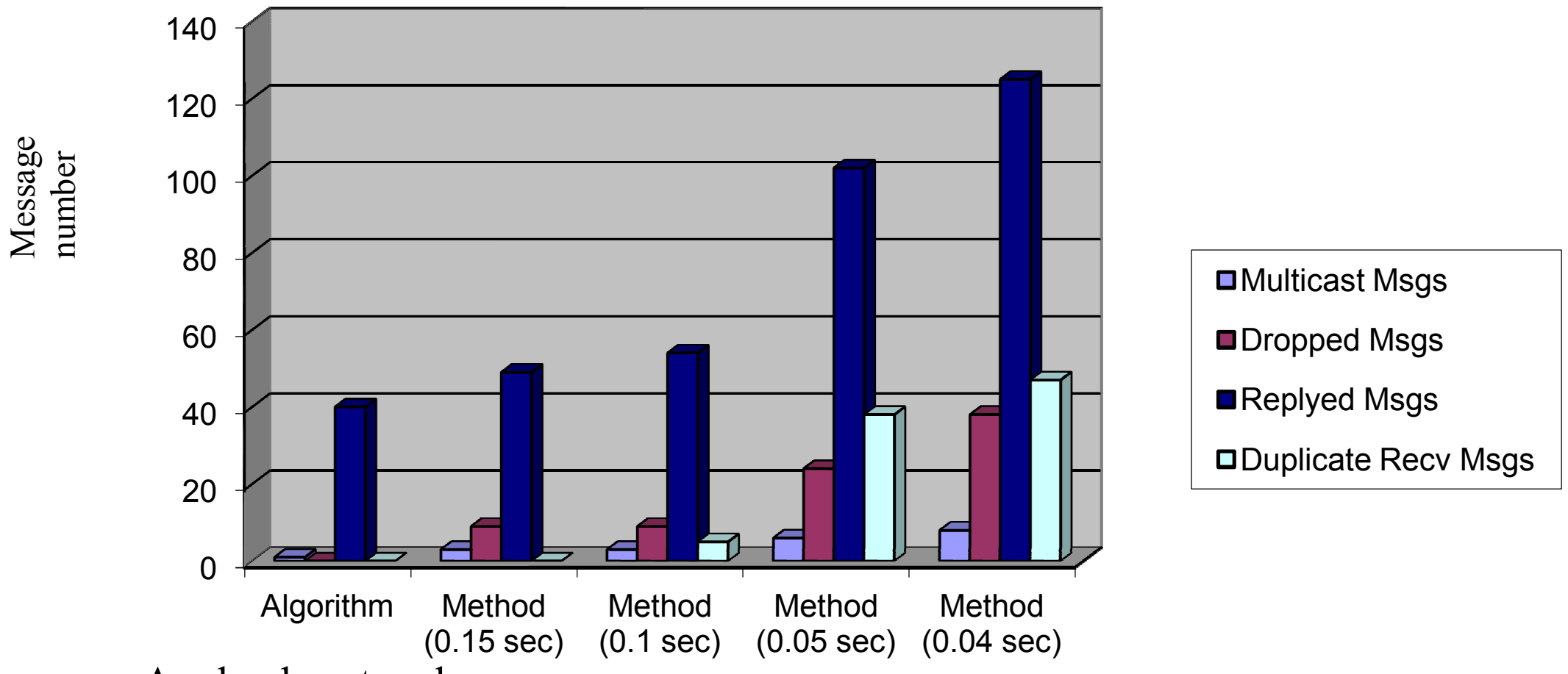

Figure 6 shows the difference in the required sent messages in the two ways

### 6.2.2 Centralized model

The network design includes 7 routers 5 of the routers are connected to 4 services (in white colour), R2 is connected to two clients (in dark colour), and R6 represent central router as shown in F 6. The applied scenario is same as decentralized model.

The proposed algorithm is used to calculate the required interval determined to be 0.048 seconds. Then the maximum limit method is used with different values (0.15, 0.1, and 0.05 seconds). Table 3, F 7 and F 8 shows the results:

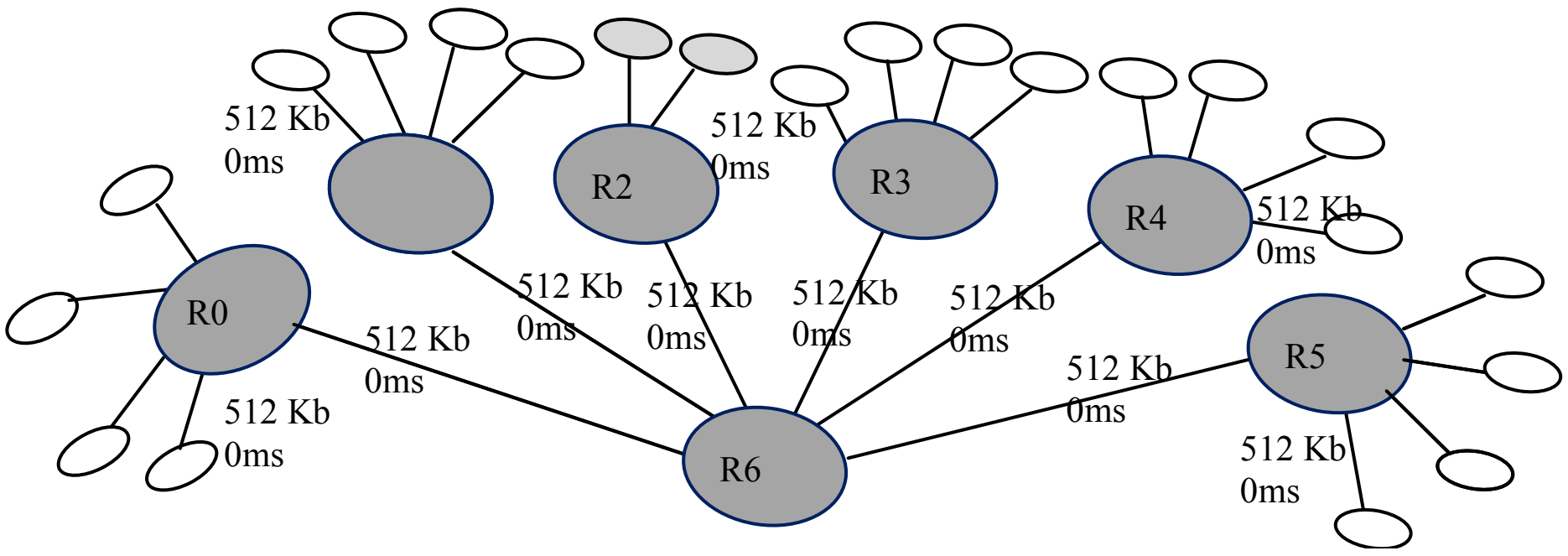

Table 3 shows a comparison between the proposed algorithm and the Maximum limit discovery method

| **Algorithm** | **Maximum limit discovery method** | | |
|---|---|---|---|
| | **0.15 second** | **0.1 second** | **0.05 second** |
| 1-Clients need to send one discovery message only (multicast messages) | 1- Clients need to send Three discovery messages (multicast messages) | 1- Clients need to send Four discovery messages (multicast messages) | 1- Clients need to send Nine discovery messages (multicast messages) |
| 2- No dropped messages | 2- There are dropped messages round 16 messages | 2- There are dropped messages round 20 messages | 2- There are dropped messages round 65 messages |
| 3- Number of the sent replying messages is 40 messages | 3- Number of the sent replying messages is 56 messages. | 3- Number of the sent replying messages is 73 messages. | 3- Number of the sent replying messages is 153 messages. |
| 3- Receiving Duplicate 0 | 3- Receiving Duplicate 0 | 3- Receiving Duplicate 13 | 3- Receiving Duplicate 44 |
| 4- Discovery process needs 0. 2 seconds to be achieved | 4- Discovery process needs 0.45 seconds to be achieved | 4- Discovery process needs 0.4 seconds to be achieved | 4- Discovery process needs 0.45 seconds to be achieved |
| 5- All services send one reply messages to each client at one time. No repeating in sending the replying messages | 5- First phase clients discovered 60 % of the services, at second phase 100% of the services discovered, third phase to make sure. | 5- First phase clients discovered 37.5 % of the services, at second phase 80%, third phase 100%, fourth phase to make sure. | 5- First 7.5 % , second 35%, third 50%, fourth 62.5%, fifth 77.5%, sixth 90%, seventh 95% eighth 100%, phase to make sure. |

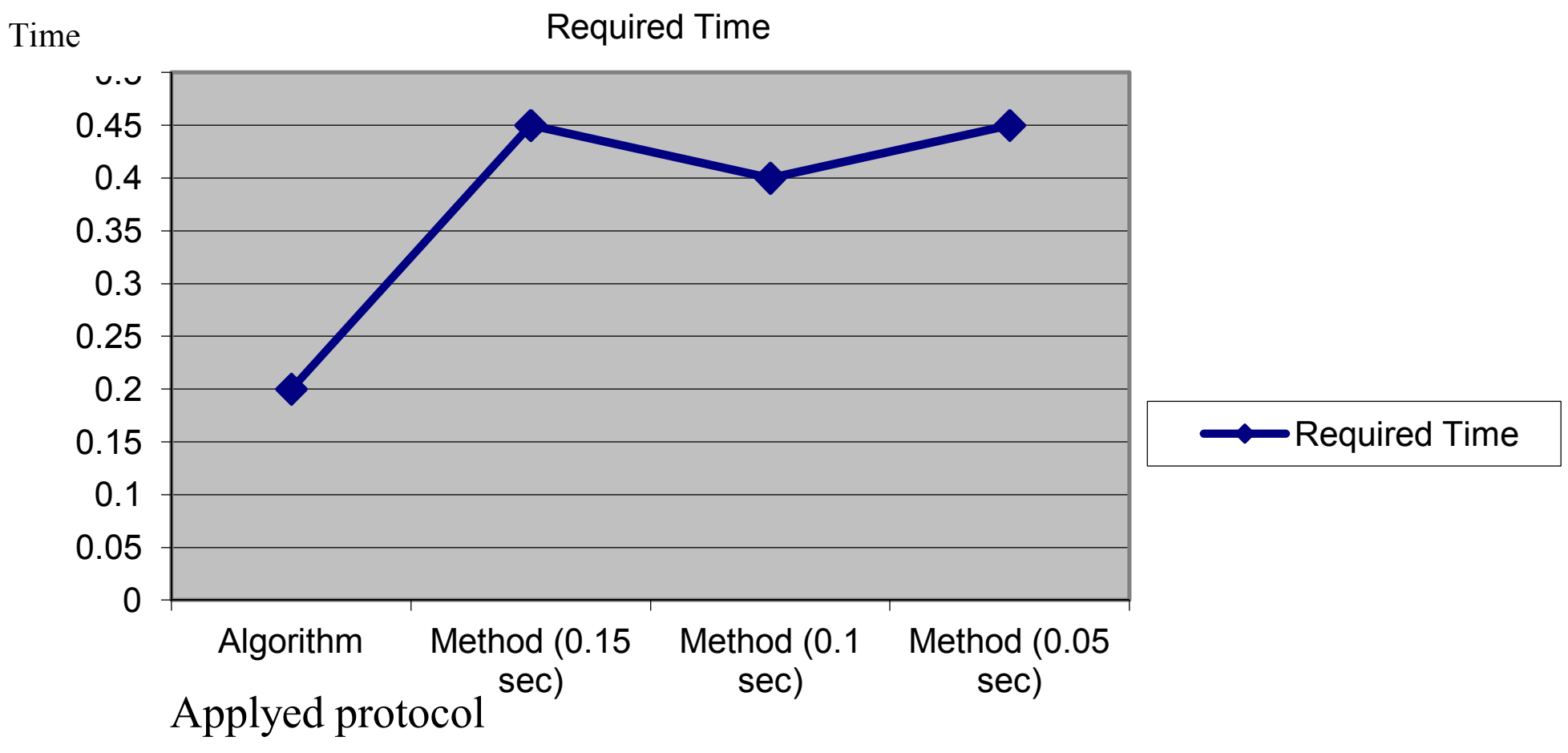

Figure 7 shows the difference in the required time in the two ways

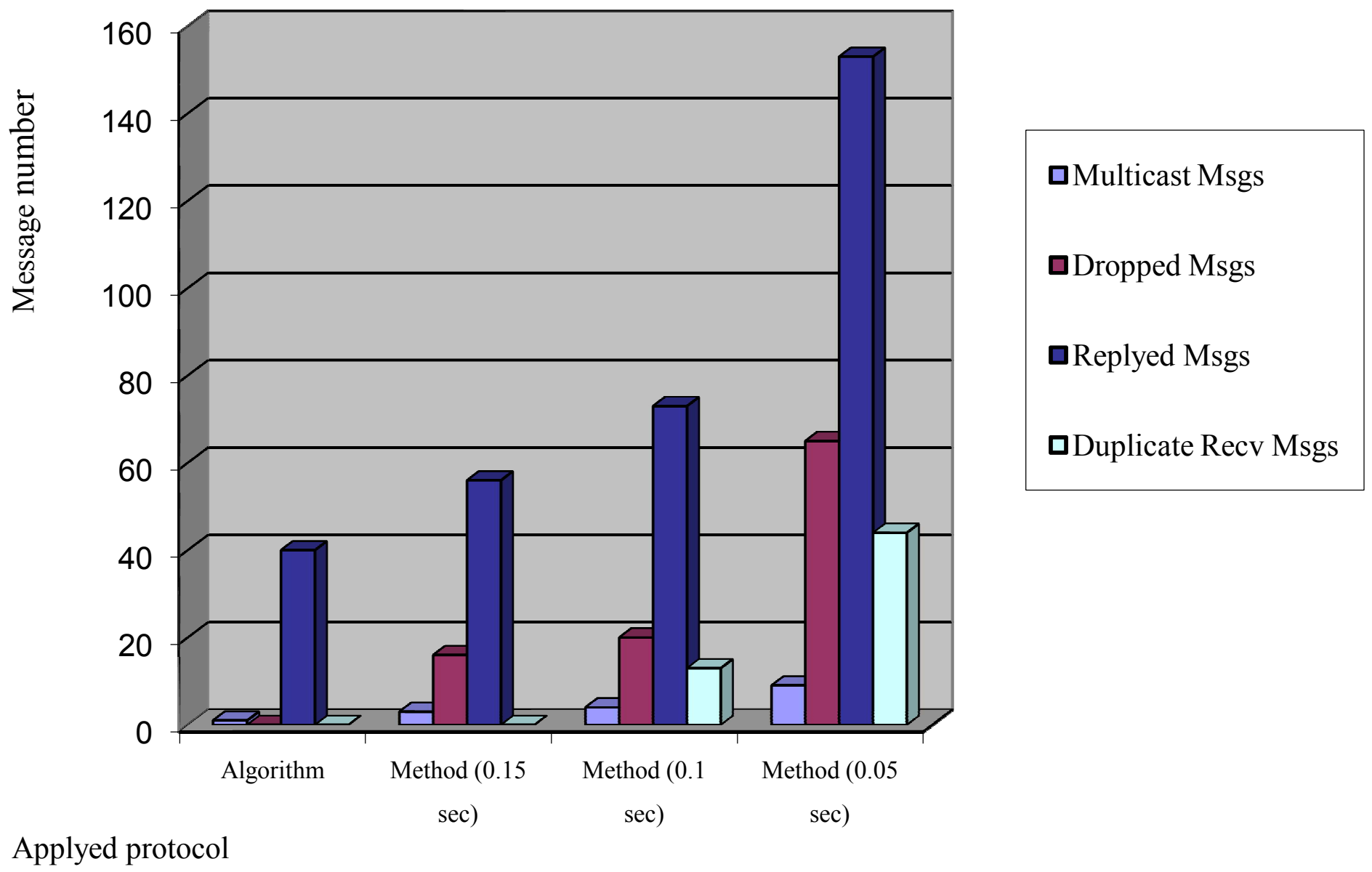

Figure 8 shows the difference in the required time in the two ways

# 7. CONCLUSION

The proposed algorithm introduces a new method in determining the relation between the required available space in receiving queue size for router with the needed interval between two consecutive burst of messages, in order to avoid losing messages. It suggests two different approaches to calculate the required receiving queue size, best interval and choosing candidate routers, depending on the network configuration (centralized or decentralized).

This algorithm takes into account the bandwidth and the message size, in other words the sending and receiving rates. The used rules in computing queue size and best interval can be modified easily in order to reduce the best interval by increasing receiving queue size and vies versa.

The results from previous experiments show no lost of messages in routers nodes and explore the influence of network configuration or design on the needed queue size and long of interval. Although scenarios one and two have the same number of clients and services, but the connection among routers different (as explained previously), the values of required receiving queue size and best interval are higher in centralized network than the decentralized network.

This is because in centralized network the central router affords all the pressure in forwarding the incoming messages which coming from all routers (except one that acts as receiving router). While in decentralized network all routers are participating in the responsibility of forwarding the incoming messages to its destination. This paper presented analyzing about the causes and location of dropping messages and explained the relation between the location of dropping messages and the used interval long.